\documentclass[slac_one]{revtex4}
\pdfoutput=1
\usepackage{graphicx}
\usepackage{fancyhdr}
\begin{document}

\title{Recent theoretical progress in perturbative QCD} 

\author{G. Zanderighi}
\affiliation{The Rudolf Peierls Centre for Theoretical Physics, 1
Keble Road, University of Oxford, Oxford.}

\begin{abstract} We review selected recent theoretical activity in
perturbative QCD. We focus on progress in the description of parton
densities, including latest developments in neural network parton
densities, on the description of high multiplicity final states at the
LHC at leading and next-to-leading order, on progress in
next-to-next-to-leading order, and on novel developments in jets
physics.
\end{abstract}

\maketitle

\thispagestyle{fancy}

\section{Introduction}
When this talk was presented, the LHC start-up was imminent. By the
time this contribution is being written first beams successfully
circulated in the LEP tunnel.\footnote{Unfortunately, also an incident
occurred, delaying further operation by a few months.} After several
years of planning, development and construction, an extremely exciting
time awaits us. The unprecedented potential of the LHC, both in terms
of energy and luminosity, will allow us to explore the region where
electroweak symmetry breaking is expected to occur.
For theorists the challenge is to provide theoretical predictions that
match or exceed the accuracy of data, aiming at an early success in
understanding this energy frontier.
We give here a brief status report and a review of selected recent
developments in perturbative QCD focusing on applications to the
LHC.\footnote{Apologies for all the important work not covered here.}

The role of QCD as the theory of strong interactions is
incontestable. We do not test QCD any longer, we use it. Despite this
it is important to keep in mind that in the last years a naive
application of well-established ideas to different contexts led
sometimes to failures and mistakes at first, but subsequently resulted
in new findings and developments. Examples are the discovery of
non-global logarithms~\cite{Dasgupta:2001sh}, which enter in
observables sensitive only to radiation in a selected phase space
region and of super-leading logarithms which point to a breakdown of
coherence~\cite{Forshaw:2006fk}. Given this past experience and the
fact that the LHC is unexplored ground with potential surprises ahead,
one needs to proceed with care. Bearing this in mind, we want to use
QCD to provide precise predictions of input parameters ($\alpha_s$,
$m_t$, parton densities\dots) and of signal to background ratios.

The prerequisite when applying perturbative QCD techniques to
high-energy scattering processes is factorization, which implies that
hadronic cross sections (differential in some ensemble of variables
called $X$ below) can be written as a convolution of partonic cross
sections $\hat \sigma$ with parton distribution functions $f(x,
\mu_F)$
\begin{equation}
\frac{d\sigma_{\rm pp \to hadrons}}{dX} =
\sum_{a,b} \int dx_1 dx_2
f_a(x_1,\mu_F)
f_b(x_2,\mu_F)
\times
\frac{d\hat \sigma_{\rm ab \to partons}(\alpha_s(\mu_R), \mu_R,
\mu_F,x_1 x_2 Q^2)}{dX}
+{\cal O}\left(\frac{\Lambda^n_{\rm QCD}}{Q^n}\right)\,.
\label{eq:fac}
\end{equation}
The main feature of eq.~(\ref{eq:fac}) is that process specific
partonic cross sections can be computed in perturbative QCD as an
expansion in the coupling constant at leading order (LO),
next-to-leading order (NLO), next-to-next-to-leading order (NNLO),
etc., while parton distribution functions are extracted from data, but
are universal and their evolution, that is the dependence on the
factorization scale is fully calculable in perturbative QCD.
Today factorization is widely accepted and we will assume it in the
following, however one should keep in mind that despite its extensive
use, with the exception of Drell-Yan production, there exists to date
no rigorous proof of factorization at hadron colliders and recently
factorization has been claimed to breakdown in dijet production
at ${\rm N^3LO}$~\cite{Collins:2007nk}.
Assuming factorization, two ingredients are required to make
predictions for the LHC: the parton distribution functions and the
partonic scattering cross sections. In the following we will discuss
recent progress in the description of both.

\section{Parton distribution functions}

\begin{figure*}[t]

\vspace{.2cm}

\begin{flushleft}
\hspace{1.3cm}
\includegraphics[width=0.35\textwidth]{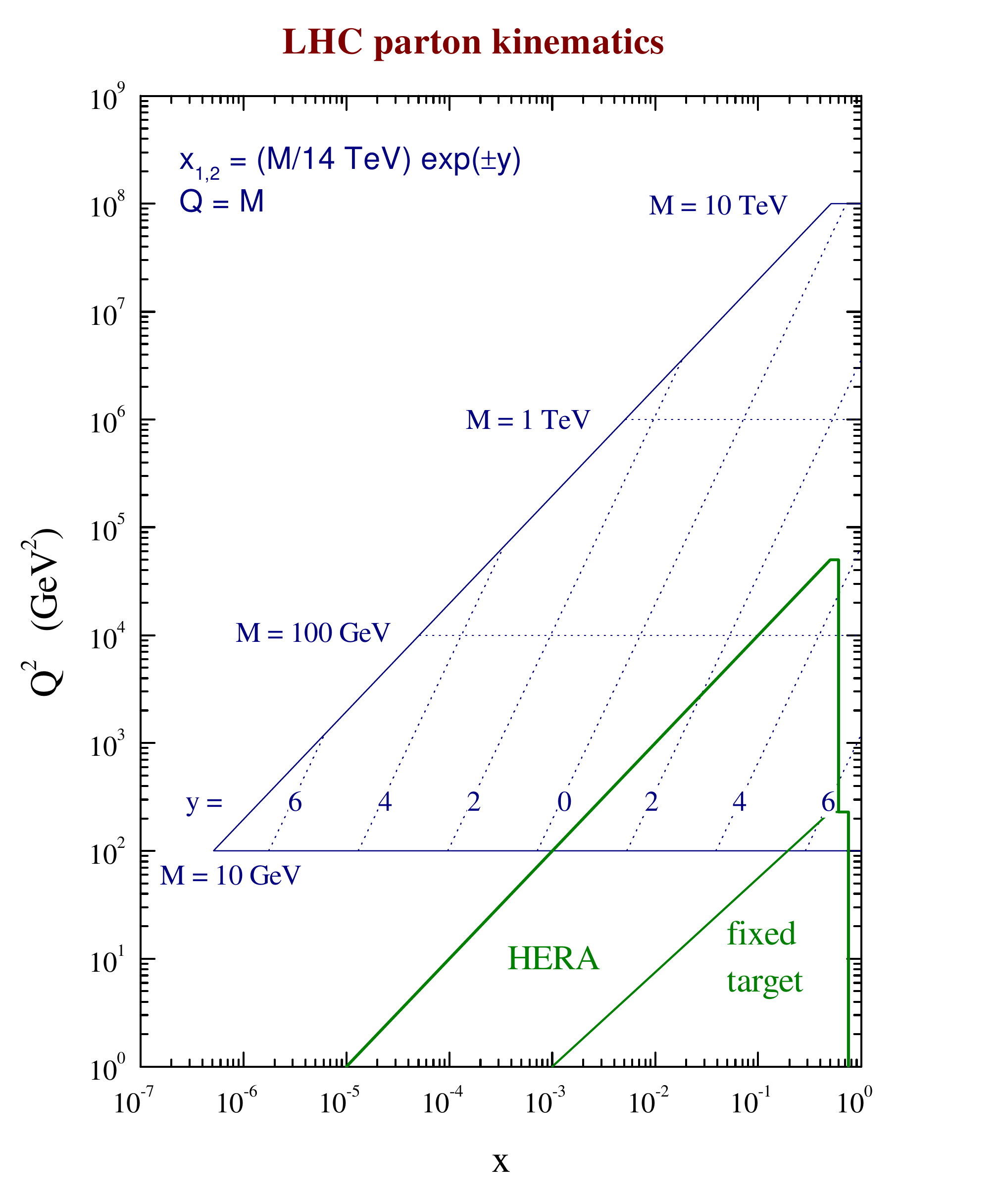}
\end{flushleft}

\begin{flushright}
\vspace{-6.cm}
\includegraphics[width=0.45\textwidth]{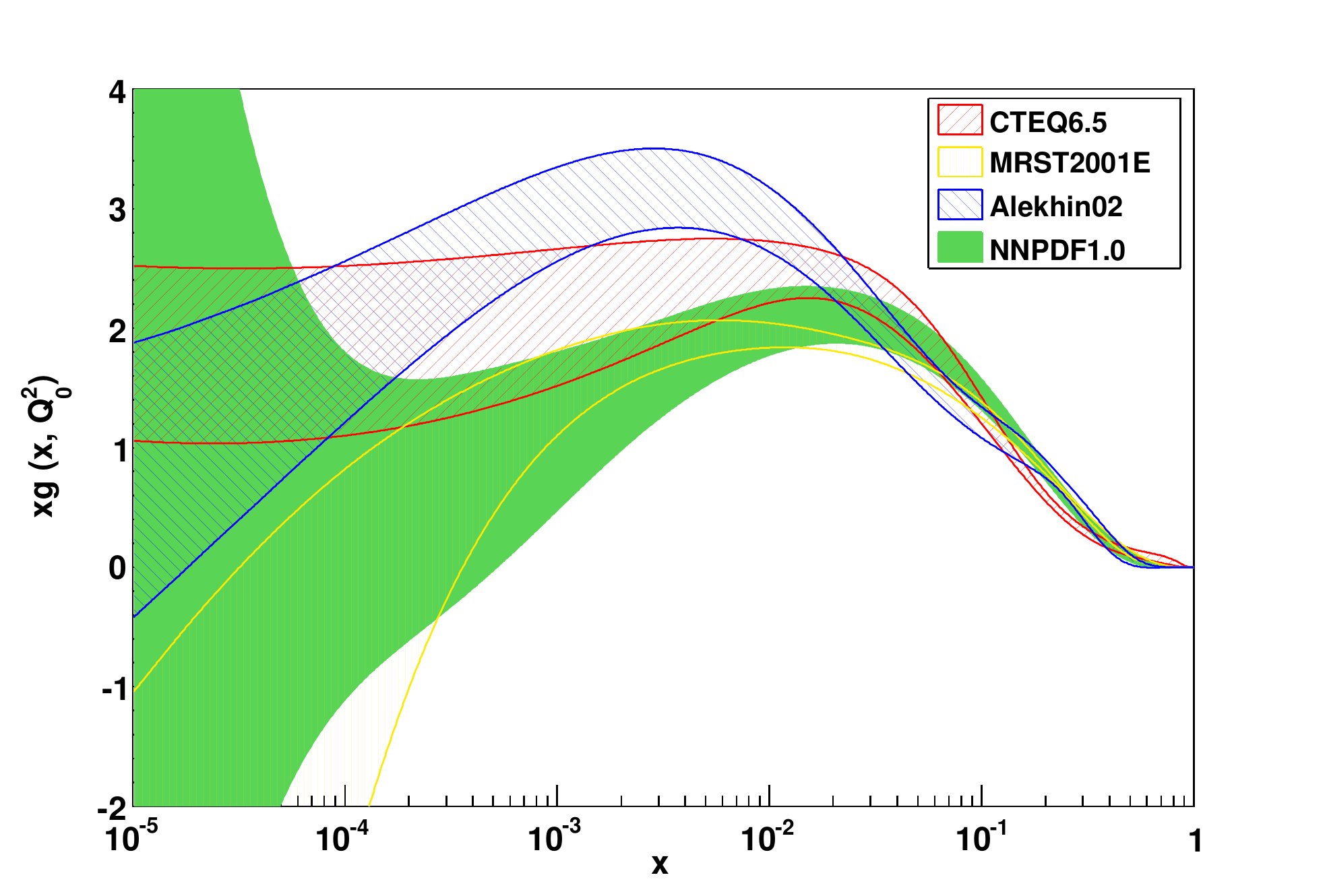}
\hspace{.5cm}
\phantom{blabla}
\end{flushright}

\caption{Left: The HERA and LHC kinematic planes (figure taken
from~\cite{Dittmar:2005ed}). Right: The gluon PDF at the starting
scale $Q_0^2= 2$ GeV$^2$ (figure taken from~\cite{NNpdfs}).}
\label{fig:pdfcov}
\end{figure*}
The most accurate extractions of parton distribution functions (PDFs)
today are from the HERA experiment.  In Fig.~\ref{fig:pdfcov} (left)
we show the parton density coverage of HERA and LHC in terms of the
energy fraction $x$ and the factorization scale $Q^2$ (the scale below
which collinear emissions are included in parton densities). The
figure illustrates that most of the LHC $x$-range is covered by the
HERA experiments, but that one needs on average a $Q^2$ evolution of 2
or 3 orders of magnitude. One also sees that while total cross
sections are mostly sensitive only to central $x$-values, rapidity
distributions probe extreme (small and large) energy fractions. Also,
while the production of states of mass of the order of $W,Z$ bosons is
mostly sensitive to small-$x$ (sea) PDFs, potentially interesting TeV
scale objects are sensitive to large $x$ values. Altogether is it
important to keep in mind the key role of HERA in providing essential
input to the LHC.

From a theoretical side recent major progress includes the NNLO
evolution of parton densities, possible because of the pillar
calculation of splitting functions at NNLO by Moch, Vermaseren, and
Vogt in 2004~\cite{3Lpdfs} and the related release of a number of
tool-kits for the NNLO DGLAP evolution of PDFs~\cite{NNLevkits}. In
recent years a treatment of heavy flavour effects near quark masses
has been accounted for, see e.g.~\cite{HQpdfs,Buttar:2008jx}. It turns
out that in some cases heavy flavour effects are large and not covered
by the previous uncertainty estimates, which were purely statistical
and did not intend to cover such effects. For example inclusion of
heavy flavours has a (6-7)\% effect on Drell-Yan cross sections at the
LHC (while the effect is negligible at the Tevatron). The reason for
these sizable corrections in this LHC standard candle process is that,
due to momentum sum rules, the presence of heavy flavours in the
initial state causes a suppression of light quarks at small $x$.

Recently Neural Network (NN) PDFs have been introduced~\cite{NNpdfs}.
These exploit standard NN techniques to provide density fits which do
not assume a specific parton density parametrization form, as is the
case for all other currently available PDFs and use replica data to
treat errors.
To understand the importance of NN PDFs it is useful to examine the
errors quoted by different groups when performing global fits.
As can be seen for example in fig.~\ref{fig:pdfcov} (right), while
various PDFs tend to agree in the central $x$ region covered by data,
they disagree in the extrapolation regions, both at very small and at
very large $x$ where different parametrization bias the distributions
in different directions. In particular, various PDFs do not agree
within the quoted uncertainties, which represent statistical errors
only. This suggests that the dominant uncertainty is due to the
specific choice of the PDF parametrization used in the fits. This
uncertainty is difficult to estimate with standard techniques, NN PDFs
on the other hand are free of this parametrization bias.
Additionally with standard PDFs it is sometimes not clear whether it
is better to focus on selected, very clean data, or whether it is
advantageous to include more and more data in the fits, even if
different data sets taken separately would give inconsistent
PDFs. With NN PDFs one naturally just includes all available data so
as to make the best out of all experimental input.
Altogether, the description of PDFs is reaching precision, but still
some work ahead is required and more progress is to be expected soon.

\section{Multiparticle final states}
The LHC will operate in a new regime with the highest energy and
luminosity ever reached. This implies that we expect to produce a very
large number of high-multiplicity events, both because typical
Standard Model (SM) processes can be accompanied by the radiation of
many jets and because most Beyond the Standard Model (BSM) signals
involve pair production and subsequent decay chains. This means that
at the LHC it will be more important to describe high-multiplicity
final states than ever before.

\paragraph{Leading order}
The simplest description in exact perturbative calculations is at
leading order. Here partons (or particles) should be well-separated
and hard so as to avoid large soft-collinear corrections. Leading
order calculations are today fully automated, and can have up to 8
particles in the final state, depending on the process under
consideration~\cite{MatrixEl}. This is a remarkable progress compared
to few years ago when the best approximation for multi-particle
production was often based on a hard $2\to2$ scattering followed by
the emission of the remaining particles in the soft-collinear
approximation, as implemented in Herwig or Pythia. The drawback of
leading order calculations is that they have very large scale
dependencies, enhanced sensitivities to kinematical cuts and a poor
modeling of the jet structure (each parton corresponding to a
jet). For example, in the simple case of the $W$+4 jets cross section
(a process relevant for many BSM searches because of the presence of 4
jets, a lepton and missing transverse energy), the LO cross section is
proportional to $\alpha^4_s(Q)$. Here $Q$ denotes an arbitrary scale,
which should be chosen to be of the order of the hard scale of the
process, but is otherwise unspecified. The residual scale dependence
of the cross section can be used to gauge the uncertainty of the
result. If one varies this scale in such a way that the coupling
varies by $\pm 10$\%, the cross section for $W$+4 jets will vary by
$\pm 40$\%. This illustrates the large uncertainties of LO
calculations. In many cases, where NLO corrections are known, the LO
result turns out to be off by ${\cal O}(100\%)$. One might therefore
wonder why we care about LO calculations at all. LO is always the
fastest option and often the only one. Today LO can be matched to
parton showers, providing a framework to test quickly new ideas in a
fully exclusive description. Also, currently there are various working
and very well-tested LO generators. Given the complexity of the
processes and observables studied at the LHC, the importance of this
should not be underestimated. To summarize, LO is highly automated,
crucial to explore new ground, but lacks precision.
It is interesting to mention which techniques beyond standard Feynman
diagrams are available at LO. One powerful method developed by Berends
and Giele (BG) 20 years ago uses off-shell currents to construct
amplitudes recursively using the building blocks given by the vertices
present in the Lagrangian~\cite{Berends:1987me}.
Britto-Cachazo-Feng (BCF) relations allow one to compute any helicity
amplitude via on-shell recursions using complex momentum
shifts~\cite{Britto:2004ap}.
Finally with Cachazo-Svrcek-Witten (CSW) relations one can compute
any helicity amplitude by sewing together MHV
amplitudes~\cite{Cachazo:2004kj}.
It is interesting to compare the numerical performance of those
methods. A study is shown in Tab.~\ref{tab:LOComp}, which shows the
time in seconds to compute $10^4$ amplitudes for $2\to n$ gluon
production. The numerical superiority of BG relations can be seen clearly
here.

\begin{table}
\begin{tabular}{|c|c|c|c|c|c|c|c|c|c|}
\hline
Final state  & 2g & 3g & 4g & 5g & 6g & 7g & 8g & 9g & 10g \\
\hline 
BG  & 0.28&   0.48&   1.04&   2.69&   7.19&   23.7&   82.1&   270 &   864 \\
BCF & 0.33  & 0.51  & 1.32  & 7.26  & 59.1  & 646   & 8690  & 127000&- \\
CSW & 0.26& 0.55& 1.75& 5.96& 30.6& 195 & 1890& 29700& -  \\
\hline 
\end{tabular}
\caption{Time in seconds to evaluate $2\to n$ gluon amplitudes for
$10^4$ points (numbers taken from~\cite{Duhr:2006iq}, see
also~\cite{Dinsdale:2006sq}).}
\label{tab:LOComp}
\end{table}

\paragraph{Next-to-leading order}
While LO is a tool to explore new ground, for precision studies NLO is
mandatory, simply because the QCD coupling is not very small, so that
NLO corrections can be numerically large. Benefits of NLO include a
reduced dependence on unphysical scales, a better modelling of jets,
and a more reliable control of the normalization and shape of cross
sections. The status of NLO corrections can be summarized as follows
(see also \cite{Bern:2008ef})
\begin{itemize}
\item all $2\to2$ scatterings are known (or easy to compute) in the SM
      and beyond;
\item very few $2\to3$ scattering processes in the SM are still not
      known;~\footnote{There is however the important caveat that many recent
      calculations do not include decays and newest codes are mostly
      private, limiting the possibility to perform realistic
      experimental studies.}
\item $2\to4$ is a hardly explored ground~\cite{Bredenstein:2008zb} and
      today no $2\to4$ LHC scattering process is fully known at NLO.
\end{itemize}
      
Three ingredients are needed to compute a $2\to N$ process at NLO
\begin{itemize}
\item the real radiation of one parton from the $2+N$ parton system
      (tree-level $2+N+1$ processes); 
\item one-loop virtual corrections to the $2\to N$ process; 
\item a method to cancel the divergences of real and virtual
      corrections before numerical integration.  
\end{itemize}

As discussed before, the calculation of tree level amplitudes has been
automated and also the cancellation of divergences is well understood
\cite{subtraction} and various automated subtraction methods based on
the Catani-Seymour dipole approach have been
formulated~\cite{subimplemented}.
Therefore at NLO the bottleneck has been the calculation of virtual,
loop amplitudes.
Because of the importance of these calculations there has been a
community effort in the last years following two main streamlines. The
first one uses traditional Feynman diagram methods supplemented by
robust numerical techniques (integration by parts, Passarino-Veltman
and Davydychev reduction, expansions or interpolations close to
numerically unstable points etc.).
Despite being very
advanced~\cite{Davydychev:1991va,Duplancic:2003tv,Giele:2004iy,Ellis:2005zh,denner,vanHameren:2005ed,Binoth:2008uq}
there computational tools 
are plagued by a factorial growth, both in the number of Feynman
diagrams to be considered and in the number of terms generated by the
tensor reduction. While these methods have been successfully applied
to some $2\to3$ processes (see e.g~\cite{Bern:2008ef}), it turns out
to be very difficult to go beyond that~\cite{Ellis:2006ss}.
Alternatively, one can exploit analytical methods, which allow one to
compute loop amplitudes without doing any loop
integration. Unfortunately, with most analytical techniques it was
possible to obtain only partial results. For instance using a
supersymmetric decomposition of QCD amplitudes one could compute
``easily'' the ${\cal N}=4$ and ${\cal N}=1$ contributions to the
amplitude, but a full QCD calculation requires also the
non-supersymmetric scalar contribution.
Also, one was able to compute one-loop gluon amplitudes with any
number of gluons in the final state, but only for specific helicity
configurations, see e.g.~\cite{Forde:2005hh}. Cross sections on the
other hand require a sum/average over all possible allowed helicity
states.
So despite their remarkable interest, only very few analytical results
ended in having a phenomenological application (important examples
are~\cite{Bern:1993mq,Bern:1997sc}). Because of the high complexity of
LHC final states, it is clear that for a modern method to be
competitive, it should allow one to compute full amplitudes and it
should be systematic, making it possible to automate the calculation
of the virtual correction, as is currently done for tree-level and for
subtraction terms.

Before sketching modern techniques for calculating one-loop
amplitudes, we present one NLO example: the recent calculation of
$t\bar t$+1 jet at the LHC~\cite{ttj}.
\begin{figure*}[t]
\centering
\includegraphics[width=6.5cm,height=4.5cm]{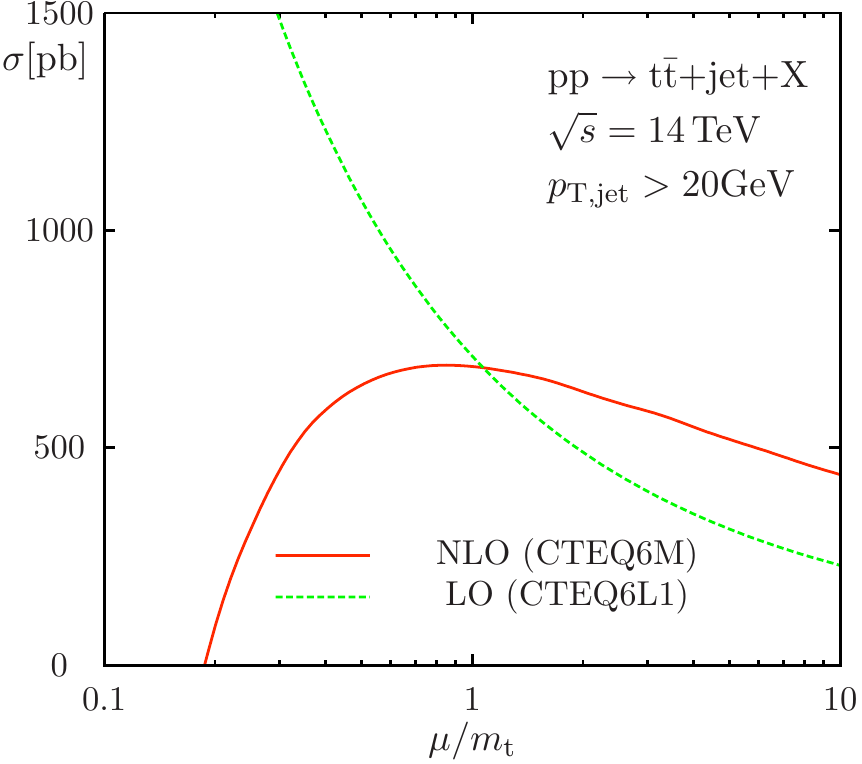}
\hspace{1cm}
\includegraphics[width=7cm,height=4.5cm]{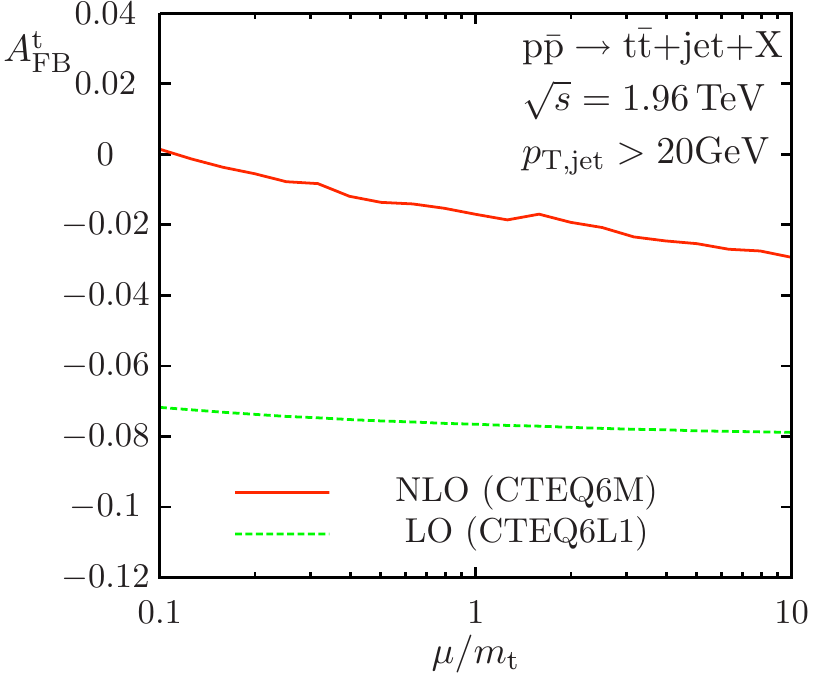}
\vspace{-0.3cm}

\caption{Left: Scale dependence at LO and NLO for the $t\bar t$ + jet
cross section at the LHC (left) and for the forward-backward charge
asymmetry for the top quark at the Tevatron (factorization and
renormalization scales are set equal). Figures taken from~\cite{ttj}.}
\label{fig:ttj}
\end{figure*}
Fig.~\ref{fig:ttj} shows the improved stability of the NLO result
(left panel) and the large effect of the NLO correction on the
forward-backward top asymmetry measured at the Tevatron, which is
compatible with zero at NLO (right panel).
It is also remarkable that the scale variation at LO is very flat and
not a good measure of the size of the NLO correction (essentially
because of cancellations between terms in the numerator and
denominator in the asymmetry). Finally we note that this calculation
constitutes an essential ingredient for the NNLO $t\bar t$ production
cross section (though an expansion to higher orders in $\epsilon$ is
needed and has recently been calculated in the case of gluons in the
initial state~\cite{Anastasiou:2008vd}). Other recently computed
calculations include Higgs+dijet production, three vector boson
production, $WW$+1jet, $t\bar t Z$, $Wb \bar b$, etc. All those
calculations have been done using traditional, Feynman diagram based
methods. For a recent review we refer the reader
to~\cite{Bern:2008ef}.

In recent years we saw a large number of novel analytical ideas, in
the following we will mention only two breakthrough ones. The first
one is the formulation of generalized unitarity ``\dots we show how to
use generalized unitarity to read off the (box) coefficients. The
generalized cuts we use are quadrupole cuts
\dots''~\cite{Britto:2004nc}.
Generalized (quadrupole) cuts in four dimensions completely freeze the
loop integration and allow one to compute the box coefficients as
products of tree-level amplitudes. The important observation is that
because the solution to the on-shell conditions correspond to complex
momenta, tree-level amplitudes with 3 on-shell gluons are non-zero
(while they would vanish if all momenta were real). The second
breakthrough idea goes now under the name of the 
Ossola-Papadopoulos-Pittau (OPP) method ``We show how to extract the
coefficients of 4-, 3-, 2-, 1-point one-loop scalar
integrals\dots''~\cite{Ossola:2006us}. The idea here is to just set up
a system of equations in the loop-momentum and reduce the problem to
an algebraic problem of finding solutions to those equations.

\begin{figure*}[t]
\begin{flushleft}
\phantom{x}
\hspace{1.cm}
\includegraphics[height=5.cm]{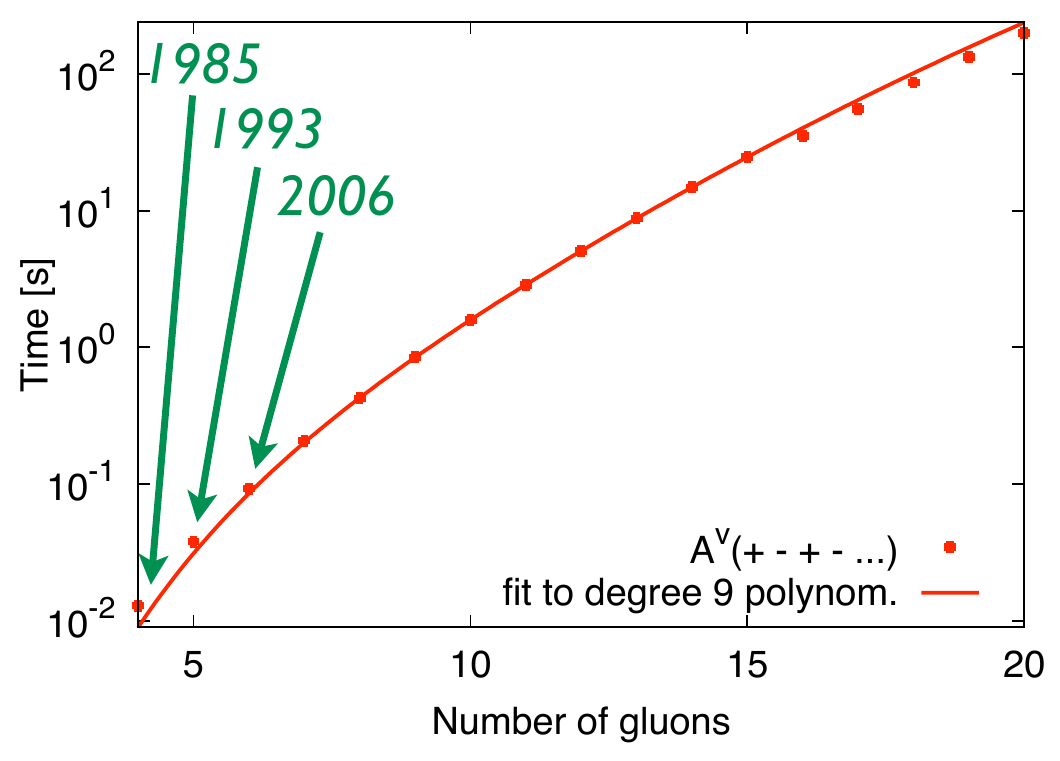}
\end{flushleft}

\begin{flushright}
\vspace{-5.5cm}
\hspace{2cm}
\includegraphics[width=7.cm,height=5.cm]{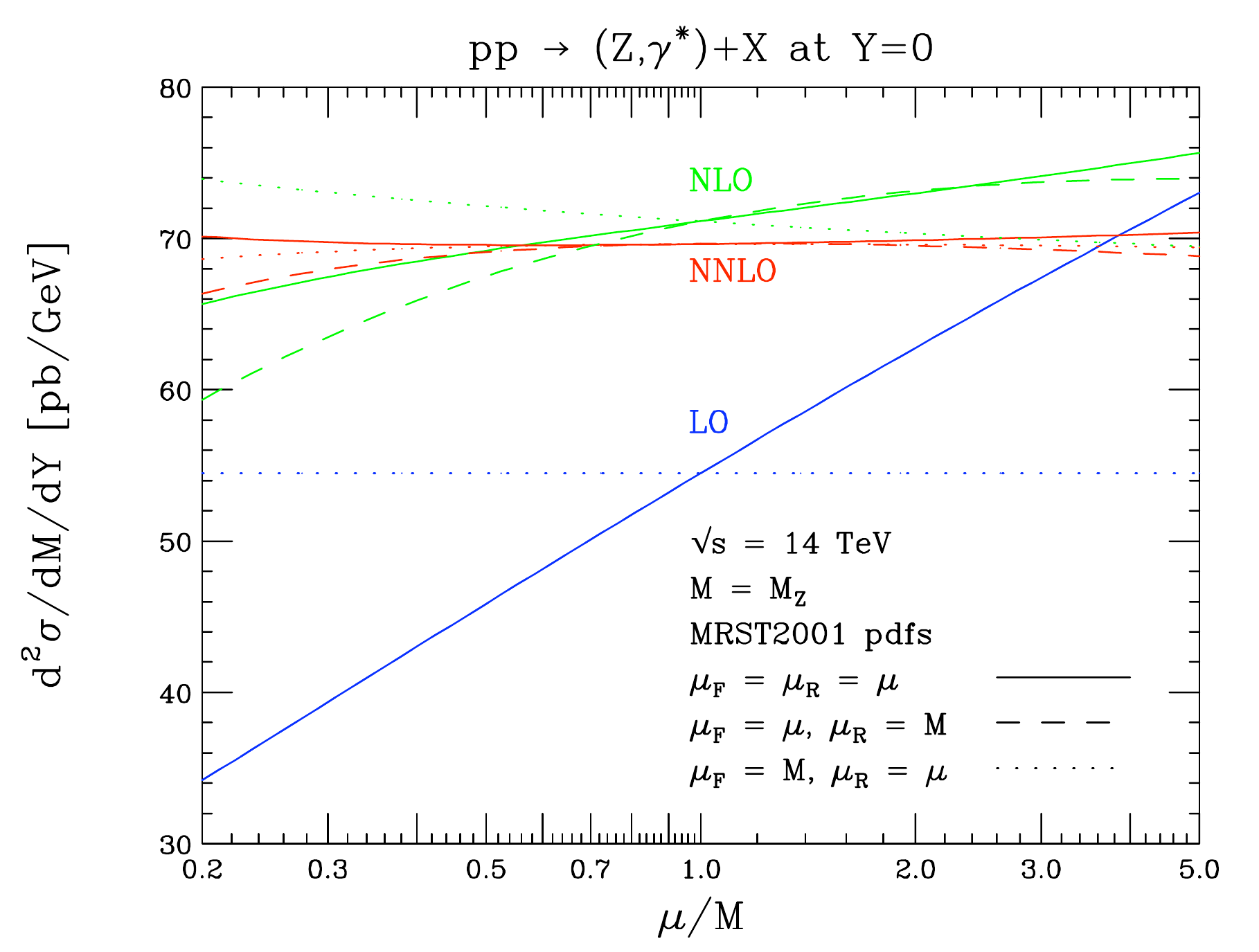}\hspace{1.cm}\phantom{blabla}
\end{flushright}
\vspace{-0.2cm}

\caption{Left: Time to compute a one-loop $N$-gluon amplitude as a
function of the number of gluons. Also shown is a fit to degree 9
polynomial (figure taken from~\cite{Giele:2008bc}).  Right: Scale
variation for the production of an on-shell $Z$ boson at the LHC at
central rapidity, $y=0$ (figure taken
from~\cite{Anastasiou:2003ds}).}
\label{fig:gluonsandDY}
\end{figure*}

A unified approach suggests merging partial fractioning via OPP with
generalized unitarity in integer higher dimensions so as to get the
cut-constructable as well as the rational
part~\cite{Giele:2008ve}. This method allows one to get full one-loop
amplitudes from tree-level amplitudes, computed
e.g. with Berends-Giele recursion relations. 
When such an approach is formulated two issues arise: the one
of practicality (speed and stability) and the one of generality,
i.e. whether the method can be used for realistic LHC processes which
involve light and heavy fermions, gluons and vector bosons.
As far as practicality is concerned, an excellent performance of the
method has been demonstrated in the case of pure gluonic
amplitudes~\cite{Giele:2008bc}. This is illustrated in
fig.~\ref{fig:gluonsandDY} (left) which shows the time needed to
evaluate the most difficult alternating sign helicity amplitudes as a
function of the number of gluons ranging between 4 and 20. Also
stability of the results is not an issue as long as one evaluates a
few unstable points in quadruple precision.
Similar results for pure gluonic amplitudes have been obtained using
the unitary method and on-shell recursions~\cite{Berger:2008sj}.
As far as applying the method to realistic LHC processes, beyond pure
gluonic amplitudes two cases have been considered now: $gg\to
ttg$~\cite{Ellis:2008ir} amplitudes and all leading and sub-leading
$0\to q\bar q W ggg$ and $0\to q\bar q W Q\bar Q g$
amplitudes~\cite{w3jets} relevant for $W+3$jet production.  Within a
similar approach based on on-shell methods, leading color $0\to q\bar
q W ggg$ amplitudes have also been computed~\cite{Berger:2008sz}.  

To conclude, while at the QCD plenary talk at ICHEP04 the statement
was made that no automation was in sight for loop amplitudes, today
automation for loop amplitudes is on the horizon and the focus in the
upcoming years will be finally on computing full cross sections. 

\paragraph{Next-to-next-to-leading order}

After discussing recent progress at NLO, the natural question which
arises is: how do we know if NLO is good enough or not?
Usually NLO is insufficient when NLO corrections are very large,
i.e. when the NLO correction is comparable to, or larger than, the LO
result. This may happen when a process involves very different scales,
so that large logarithms of the ratio of the two scales arise which
need to be resummed. This may also happen when new channels open up
(at NLO those channels are effectively LO), this is the case for
instance for $b$-jet production, where gluon splitting and flavour
excitation processes enter at NLO and are enhanced by large
logarithms.  Also, gluon dominated processes are often characterized
by large corrections, both because gluons radiate on average more than
quarks and because of the steeply falling PDFs at small x.
NLO might also not be sufficient if very high precision is useful,
this is occasionally the case, for instance for Drell-Yan, top pair
production, and 3-jet production in $e^+e^-$.
Finally, since NLO provides a first reliable estimate of cross
sections, only NNLO can in principle provide a reliable error estimate
of those cross sections.
The bottleneck at NNLO is not the calculation of virtual matrix
elements, as is the case at NLO, but rather the cancellation of
divergences before numerical evaluation.

Today Drell-Yan is the best known process at a hadron collider and
constitutes the most important and most precise test of the SM at the
LHC. Indeed Drell-Yan is known at NNLO including spin-correlation
effects, finite-width effects, $\gamma-Z$ interference and the
calculation is fully differential in the lepton
momenta~\cite{Anastasiou:2003ds,Melnikov:2006di}. Sample results are
shown in fig.~\ref{fig:gluonsandDY} which demonstrate the improved
scale stability at higher orders. 

Inclusive Higgs production at NNLO has also been known for a few years
now~\cite{Harlander:2002wh,Ravindran:2003um,Anastasiou:2004xq}. Recently, the decays of
the Higgs into 4 charged leptons or into 2 charged leptons and two
neutrinos have been
included~\cite{Anastasiou:2007mz,Anastasiou:2008ik,Grazzini:2008tf}.
This gives one the possibility to study cross sections with realistic
cuts on the jets and on the lepton momenta. It turns out that a veto
on jets with large transverse momentum dramatically decreases the size
of the NNLO corrections and improves the convergence of the
perturbative expansion.
Very recently, a study suggests that the large perturbative
corrections to Higgs production stem from $(C_A\pi\alpha_s)^n$
enhanced terms, which arise in the analytic continuation of the gluon
form factor to time-like region~\cite{Ahrens:2008qu}.

Another NNLO calculation completed after several years in 2007 is the
one of 3-jet production in $e^+e^-$~\cite{GehrmannDe
Ridder:2007bj}. The main motivation for this calculation was that the
error on $\alpha_s$ from jet-observables was dominated by the
theoretical uncertainty, $\alpha_s(M_Z) = 0.121 \pm
0.001 ({\rm exp.}) \pm 0.005 ({\rm th.})$. Higher orders were therefore
mandatory to reduce this error. 
\begin{figure*}[b]
\centering
\includegraphics[width=6cm,angle=270]{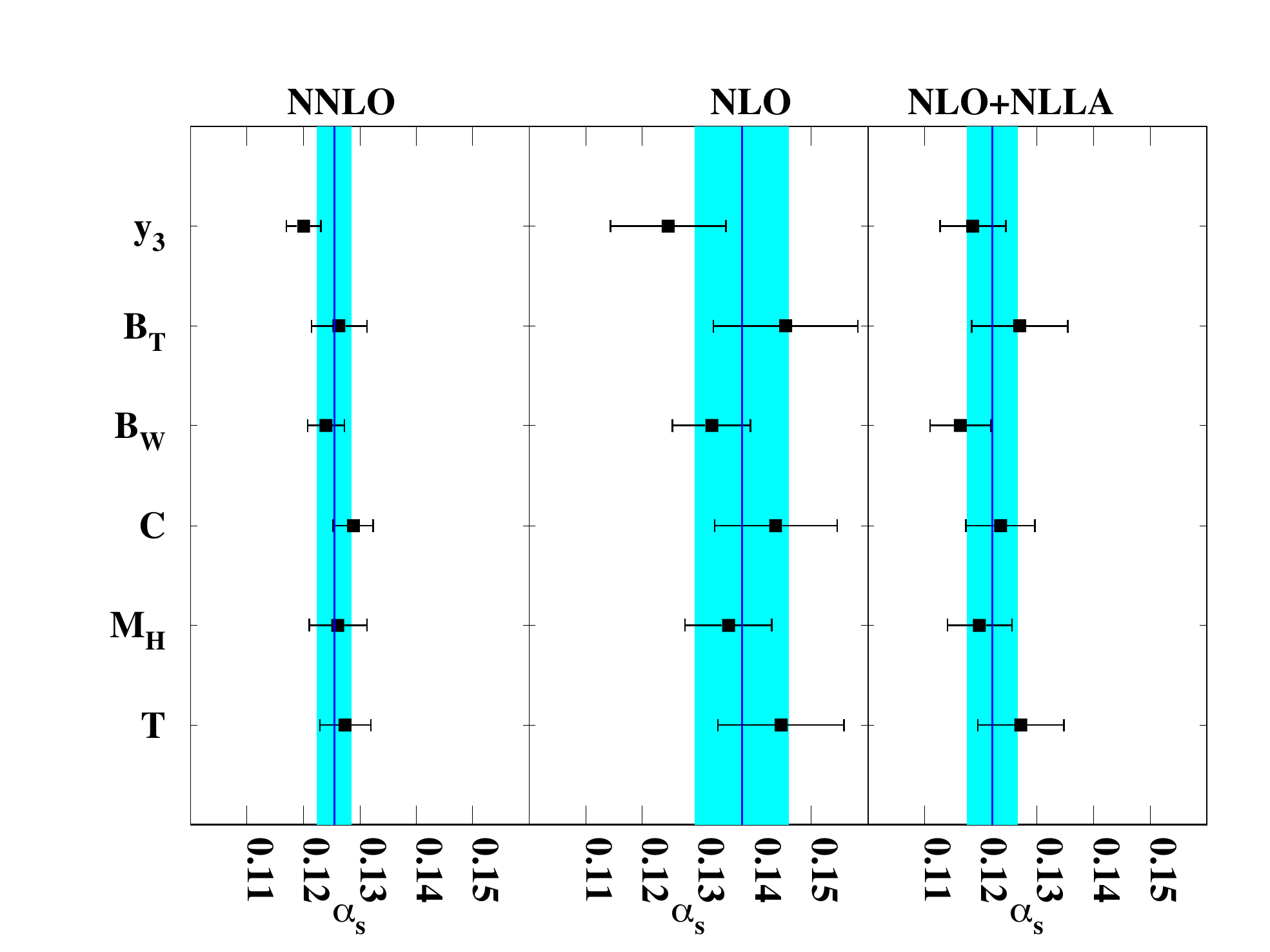}
\vspace{-0.3cm}

\caption{Scatter in the extraction of $\alpha_s(M_Z)$ from six event
shapes when using NNLO, NLO and NLO+NLL approximations (figure taken
from~\cite{Dissertori:2007xa}).}
\label{fig:asmz}
\end{figure*}
For this calculation a new method based on antenna subtraction at NNLO
was developed. The first application was a NNLO fit of $\alpha_s$ from
event-shapes. As illustrated in fig.~\ref{fig:asmz} the scale
variation effect is reduced at NNLO by a factor of 2, the scatter
between $\alpha_s$ from different event shapes is also reduced, and
the fit gives a central value closer to the world average with a
better $\chi^2$ value.

More recently, this NNLO calculation has been used in conjunction with
a ${\rm N^3LL}$ resummation for the thrust distribution using soft
collinear effective theory (SCET) methods~\cite{Becher:2008cf}. This
impressive calculation gives an $\alpha_s$ which lies on top of the
world average with an error comparable to that of the world
average. However, SCET resummation points to potential problems in
some (2 or 3) color structures in the low thrust region.
After that, an independent NNLO calculation confirmed the
disagreements on those color
structures~\cite{Weinzierl:2008iv}. Whether the value of $\alpha_s$
from NNLO fits will remain largely unaffected is something we will
know soon.

Other very accurate recent determinations of the strong coupling come
from $Z$ and $\tau$ decays. These calculations use cutting-edge
five-loop results for massless propagator integrals, fixed order
versus contour-improved perturbation theory and sum
rules~\cite{otheras}.
As a general feature, higher order terms lead to a stabilization of
the perturbative series, to a reduction of the theoretical uncertainly
and to a small shift in the central value. A good agreement between
the values of $\alpha_s$ from $Z$ and $\tau$ decays occurring at very
different energies is found, however comparing with lattice
simulations, partially conflicting results emerge.

\section{Jets}
Two years ago when discussing or reading about jets, one encountered
various qualitative statements whose origin and level of correctness
was largely unclear. Since then a tremendous progress in the
description of jets has been achieved. This includes a fast
implementation of the $k_t$ algorithm
(FastJet)~\cite{Cacciari:2005hq}, an infrared safe definition of jet
flavour and related accurate predictions for $b$-jets~\cite{jetflav},
an infrared safe cone algorithm (SISCone)~\cite{Salam:2007xv}, a new
anti-$k_t$ algorithm~\cite{Cacciari:2008gp}, new concepts of jet-areas
and related techniques for pile-up subtraction~\cite{area}, quality
measures to quantify systematically the performance of
jet-algorithms~\cite{Cacciari:2008gd}, analytical and Monte Carlo
based studies of non-perturbative effects in
jets~\cite{Dasgupta:2007wa} and studies of the jet-substructure as a
way to enhance important search channels~\cite{Butterworth:2008iy}.
In the following we will illustrate just two recent developments, for
a recent review on the subject see~\cite{Buttar:2008jx}.

\begin{figure*}[t]
\centering
\includegraphics[width=0.50\textwidth]{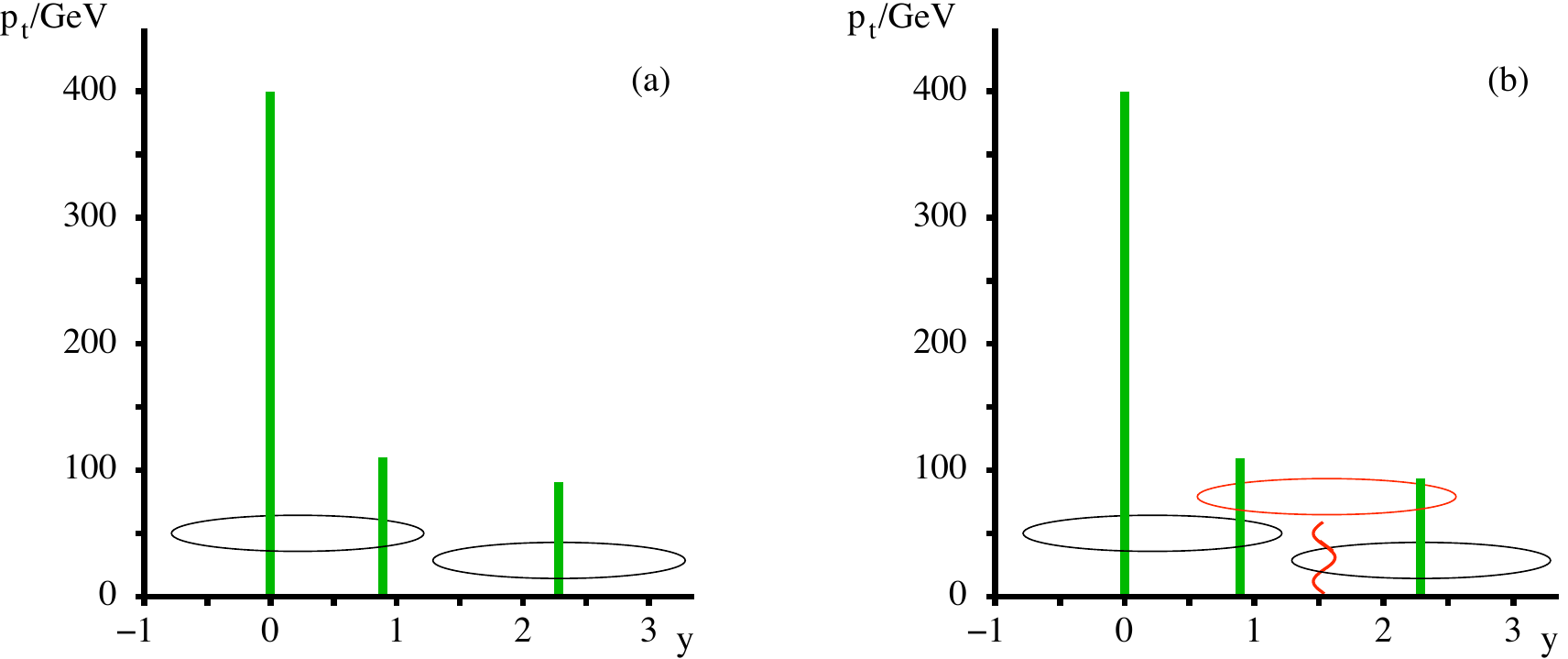}
\vspace{-0.3cm}

\caption{Illustration of IR unsafety of midpoint algorithm: a) stable
cones found with 3 hard emissions; b) stable cones found with the same
hard emission and the addition of an arbitrarily soft particle (figure taken
from~\cite{Salam:2007xv}).}
\label{fig:cone}
\end{figure*}

The first one relates to the issue of infrared
unsafety of standard, seeded cone algorithms. To understand what the
problem is it is useful to examine fig.~\ref{fig:cone} (left), which
shows schematically a simple event with 3 hard emissions. By running,
for instance, a midpoint algorithm with a given $R$ one would identify
two stable cones. When an additional very soft emission is added to
the event fig.~\ref{fig:cone} (right), the same algorithm finds 3
stable cones. This is because the additional emission provides an
additional seeds (i.e. trial direction). So the stable cone which was
missed before is now found. Because a soft emission can change the
structure of the final state hard jets the algorithm is infrared
unsafe. The solution is to use SISCone~\cite{Salam:2007xv}, a
seedless, fast ($N^2\ln N$) algorithm which finds all stable
cones. Similarly one can show that the iterative cone is collinear
unsafe and can be replaced by the anti-$k_t$ algorithm (which, just
like the iterative cone, has the property that very soft emissions are
recombined together only later on, while hard emissions eat up
efficiently everything which lies close by, hence the nice circular
shape of the jets).

One might wonder whether infrared unsafety is just an aesthetic,
formal issue with negligible practical consequences or whether it
can have a physical impact at the LHC. Some studies at the Tevatron
suggested that the difference between running an infrared safe or
unsafe algorithm on inclusive jet cross sections is less than 1\%. For
more exclusive quantities on the other hand larger differences can
occur. For instance one can find up to 40\% differences between
SISCone and midpoint when looking at the mass-spectrum of the second
hardest jet.
In general the more particles are present in the hard event, the
earlier in the perturbative expansion one will start missing stable
cones. For instance in the case of W/V/H + 2-jet cross section, cones
are missed at NLO, therefore the last meaningful order is LO (while a
NLO calculation is available in MCFM). Similarly, for jet masses in 3
jet events, cones are already missed at leading order, so that no
perturbative order is meaningful. Therefore, if one does not want
theoretical efforts to be wasted one should stick to infrared safe
algorithms.

\begin{figure*}[t]
\centering
\includegraphics[width=0.3\textwidth]{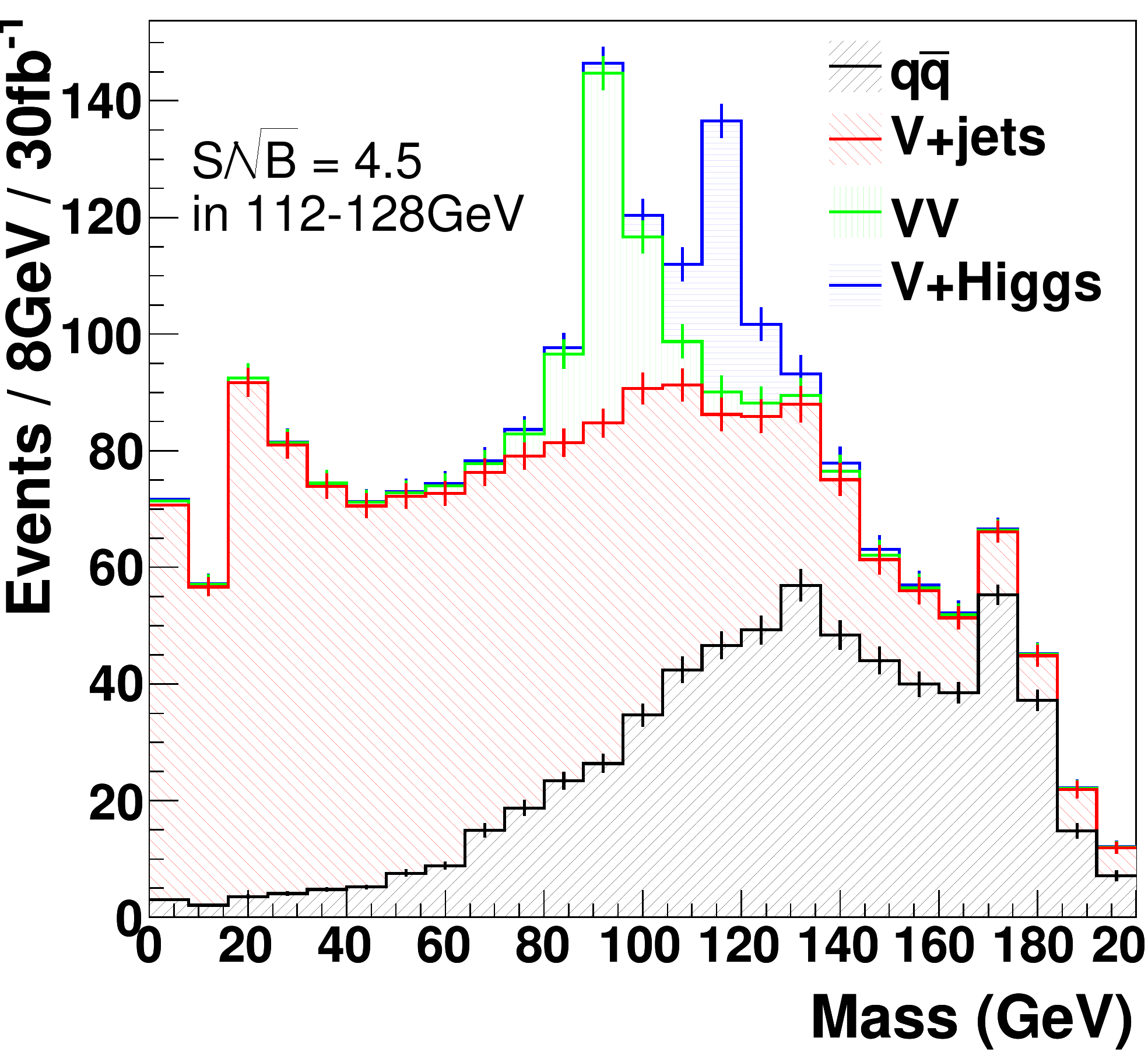}
\vspace{-0.4cm}

\caption{Signal and background for a 115 GeV SM Higgs simulated using
Herwig. The jet-clustering is C/A MD-F with $R=1.2$ and $p_T >200$
GeV. A $b$-tagging efficiency of 60\% and a mistag of 2\% is
used. $q\bar q$ denotes both heavy and light dijets. The errors
reflect the statistical uncertainty on $30$ fb$^{-1}$ (figure kindly
provided by Adam Davison).}
\label{fig:jetsub}
\end{figure*}

The second recent jet development we will illustrate here is the use
of jet substructure in the case of Higgs production. As is well-known
a light Higgs is difficult to find experimentally at the LHC and
several channels with individually low significance have to be
combined to achieve an acceptable total sensitivity. If the Higgs is
light it decays predominantly in $b\bar b$. The associated $Z/W$+
Higgs production with $H\to b \bar b$ could seem at first sight a good
channel, however, because of very large QCD backgrounds, the ATLAS
Technical Design Report concluded that ``The extraction of a signal
from $H\to b \bar b$ decays in the $WH$ channel will be very difficult
at the LHC even under the most optimistic assumptions
\dots''~\cite{AtlasTDR}. Since then this channel has been largely
neglected. Recently on the other hand, it has been recognized that
despite the loss in statistics, there are important benefits in
considering highly boosted, high $p_t$ Higgs bosons: their central decay
products give rise to a single massive
jet~\cite{Butterworth:2008iy}. One can then use a jet-finding geared
to identify the characteristic structure of the fast-moving Higgs that
decays into a $b\bar b$ pair close in angle. Concretely, one can use a
Cambridge/Aachen algorithm with a quite large radius $R$, one can then undo the
last recombination step and require that this gives rise to a large
mass drop and to a symmetric event with two $b$-tags. Finally one can
filter away the underlying event by taking only the 3 hardest sub-jets
in the event. The invariant mass of those is plotted in
fig.~\ref{fig:jetsub} for $m_H=115$ GeV when imposing standard search
cuts, combining the three channels $W\to l
\nu $, $Z\to l^+l^-$ and $Z\to \nu \bar \nu$ and assuming a real and fake
$b$-tag rates of 0.6 and 0.02. One can clearly see the Higgs peak at $115$ GeV.
Additionally, one can see a very neat peak at the $Z$ mass due to
$WZ(Z\to b\bar b$) which is important for calibration. A preliminary
study indicates that a 4.5$\sigma$ sensitivity can be obtained with
30 fb$^{-1}$. This would mean that $VH$ with $H\to b \bar b$ is
recovered as one of the best discovery channels for a light Higgs. Of
course more experimental studies are to come.

\section{Conclusions}
In the last years we have seen an impressive amount of progress in
perturbative QCD in the description of parton densities, in higher
order calculations (LO, NLO, NNLO and resummations) and in the
description of jets, where an amazing level of sophistication has been
reached.
Progress was mainly driven by automation, flexibility, release of
public codes, many new ideas and very good communication with
experimentalists, leading to several common papers.
Of course there are still many challenges ahead and we don't know yet
what we might discover at the LHC, but QCD theory will provide solid
basis for a successful physics programme at the LHC.

\begin{acknowledgments} Many thanks to Babis Anastasiou, Andrea Banfi,
Matteo Cacciari, Keith Ellis, Fabio Maltoni, Kirill Melnikov, Juan
Rojo, Gavin Salam and Gregory Soyez for valuable discussions and to
the organizers for the making ICHEP08 such an enjoyable and
stimulating meeting. The author is supported by the British Science
and Technology Facilities Council (STFC).

\end{acknowledgments}

\end{document}